
\input fontch.tex

%
%
\def\unredoffs{} \def\redoffs{\voffset=-.31truein\hoffset=-.48truein}
\def\speclscape{}
%
%
%
%
%
\newbox\leftpage \newdimen\fullhsize \newdimen\hstitle \newdimen\hsbody
\tolerance=1000\hfuzz=2pt
\catcode`\@=11 
\ifx\hyperdef\UNd@FiNeD\def\hyperdef#1#2#3#4{#4}\def\hyperref#1#2#3#4{#4}\fi
\def\bigans{b }
\def\answ{b }
%
\ifx\answ\bigans\message{(This will come out unreduced.}
\magnification=1200\unredoffs\baselineskip=16pt plus 2pt minus 1pt
\hsbody=\hsize \hstitle=\hsize 
\else\message{(This will be reduced.} \let\l@r=L
\magnification=1000\baselineskip=16pt plus 2pt minus 1pt \vsize=7truein
\redoffs \hstitle=8truein\hsbody=4.75truein\fullhsize=10truein\hsize=\hsbody
\output={\ifnum\pageno=0 
  \shipout\vbox{\speclscape{\hsize\fullhsize\makeheadline}
    \hbox to \fullhsize{\hfill\pagebody\hfill}}\advancepageno
  \else
  \almostshipout{\leftline{\vbox{\pagebody\makefootline}}}\advancepageno
  \fi}
\def\almostshipout#1{\if L\l@r \count1=1 \message{[\the\count0.\the\count1]}
      \global\setbox\leftpage=#1 \global\let\l@r=R
 \else \count1=2
  \shipout\vbox{\speclscape{\hsize\fullhsize\makeheadline}
      \hbox to\fullhsize{\box\leftpage\hfil#1}}  \global\let\l@r=L\fi}
\fi
%
\newcount\yearltd\yearltd=\year\advance\yearltd by -2000

\def\Title#1#2{\nopagenumbers\abstractfont\hsize=\hstitle\rightline{#1}%
\vskip 1in\centerline{\titlefont #2}\abstractfont\vskip .5in\pageno=0}
\def\Date#1{\vfill\leftline{#1}\tenpoint\supereject\global\hsize=\hsbody%
\footline={\hss\tenrm\hyperdef\hypernoname{page}\folio\folio\hss}}%
%

\def\draftmode{\message{ DRAFTMODE }\def\draftdate{{\rm preliminary draft:
\number\month/\number\day/\number\yearltd\ \ \hourmin}}%
\headline={\hfil\draftdate}\writelabels\baselineskip=20pt plus 2pt minus 2pt
 {\count255=\time\divide\count255 by 60 \xdef\hourmin{\number\count255}
  \multiply\count255 by-60\advance\count255 by\time
  \xdef\hourmin{\hourmin:\ifnum\count255<10 0\fi\the\count255}}}
\def\nolabels{\def\wrlabeL##1{}\def\eqlabeL##1{}\def\reflabeL##1{}}
\def\writelabels{\def\wrlabeL##1{\leavevmode\vadjust{\rlap{\smash%
{\line{{\escapechar=` \hfill\rlap{\sevenrm\hskip.03in\string##1}}}}}}}%
\def\eqlabeL##1{{\escapechar-1\rlap{\sevenrm\hskip.05in\string##1}}}%
\def\reflabeL##1{\noexpand\llap{\noexpand\sevenrm\string\string\string##1}}}
\nolabels
%
\global\newcount\secno \global\secno=0
\global\newcount\meqno \global\meqno=1
\def\s@csym{}
\def\newsec#1{\global\advance\secno by1%
{\toks0{#1}\message{(\the\secno. \the\toks0)}}%
\global\subsecno=0\eqnres@t\let\s@csym\secsym\xdef\secn@m{\the\secno}\noindent
{\bf\hyperdef\hypernoname{section}{\the\secno}{\the\secno.} #1}%
\writetoca{{\string\hyperref{}{section}{\the\secno}{\the\secno.}} {#1}}%
\par\nobreak\medskip\nobreak}
\def\eqnres@t{\xdef\secsym{\the\secno.}\global\meqno=1\bigbreak\bigskip}
\def\sequentialequations{\def\eqnres@t{\bigbreak}}\xdef\secsym{}
\global\newcount\subsecno \global\subsecno=0
\def\subsec#1{\global\advance\subsecno by1%
{\toks0{#1}\message{(\s@csym\the\subsecno. \the\toks0)}}%
\ifnum\lastpenalty>9000\else\bigbreak\fi
\noindent{\it\hyperdef\hypernoname{subsection}{\secn@m.\the\subsecno}%
{\secn@m.\the\subsecno.} #1}\writetoca{\string\quad
{\string\hyperref{}{subsection}{\secn@m.\the\subsecno}{\secn@m.\the\subsecno.}}
{#1}}\par\nobreak\medskip\nobreak}
\def\appendix#1#2{\global\meqno=1\global\subsecno=0\xdef\secsym{\hbox{#1.}}%
\bigbreak\bigskip\noindent{\bf Appendix \hyperdef\hypernoname{appendix}{#1}%
{#1.} #2}{\toks0{(#1. #2)}\message{\the\toks0}}%
\xdef\s@csym{#1.}\xdef\secn@m{#1}%
\writetoca{\string\hyperref{}{appendix}{#1}{Appendix {#1.}} {#2}}%
\par\nobreak\medskip\nobreak}
%
%
\def\checkm@de#1#2{\ifmmode{\def\f@rst##1{##1}\hyperdef\hypernoname{equation}%
{#1}{#2}}\else\hyperref{}{equation}{#1}{#2}\fi}
\def\eqnn#1{\DefWarn#1\xdef #1{(\noexpand\relax\noexpand\checkm@de%
{\s@csym\the\meqno}{\secsym\the\meqno})}%
\wrlabeL#1\writedef{#1\leftbracket#1}\global\advance\meqno by1}
\def\f@rst#1{\c@t#1a\em@ark}\def\c@t#1#2\em@ark{#1}
\def\eqna#1{\DefWarn#1\wrlabeL{#1$\{\}$}%
\xdef #1##1{(\noexpand\relax\noexpand\checkm@de%
{\s@csym\the\meqno\noexpand\f@rst{##1}}{\hbox{$\secsym\the\meqno##1$}})}
\writedef{#1\numbersign1\leftbracket#1{\numbersign1}}\global\advance\meqno by1}
\def\eqn#1#2{\DefWarn#1%
\xdef #1{(\noexpand\hyperref{}{equation}{\s@csym\the\meqno}%
{\secsym\the\meqno})}$$#2\eqno(\hyperdef\hypernoname{equation}%
{\s@csym\the\meqno}{\secsym\the\meqno})\eqlabeL#1$$%
\writedef{#1\leftbracket#1}\global\advance\meqno by1}
\def\xeqn{\expandafter\xe@n}\def\xe@n(#1){#1}
\def\xeqna#1{\expandafter\xe@n#1}
\def\eqns#1{(\e@ns #1{\hbox{}})}
\def\e@ns#1{\ifx\UNd@FiNeD#1\message{eqnlabel \string#1 is undefined.}%
\xdef#1{(?.?)}\fi{\let\hyperref=\relax\xdef\next{#1}}%
\ifx\next\em@rk\def\next{}\else%
\ifx\next#1\xeqn#1\else\def\n@xt{#1}\ifx\n@xt\next#1\else\xeqna#1\fi
\fi\let\next=\e@ns\fi\next}

\def\DefWarn#1{\ifx\UNd@FiNeD#1\else
\immediate\write16{*** WARNING: the label \string#1 is already defined ***}\fi}
%
\newskip\footskip\footskip14pt plus 1pt minus 1pt 
\def\footnotefont{\ninepoint}\def\f@t#1{\footnotefont #1\@foot}
\def\f@@t{\baselineskip\footskip\bgroup\footnotefont\aftergroup\@foot\let\next}
\setbox\strutbox=\hbox{\vrule height9.5pt depth4.5pt width0pt}
\global\newcount\ftno \global\ftno=0
\def\foot{\global\advance\ftno by1\def\foot@rg{\hyperref{}{footnote}%
{\the\ftno}{\the\ftno}\xdef\foot@rg{\noexpand\hyperdef\noexpand\hypernoname%
{footnote}{\the\ftno}{\the\ftno}}}\footnote{$^{\foot@rg}$}}
%
\newwrite\ftfile
\def\footend{\def\foot{\global\advance\ftno by1\chardef\wfile=\ftfile
\hyperref{}{footnote}{\the\ftno}{$^{\the\ftno}$}%
\ifnum\ftno=1\immediate\openout\ftfile=\jobname.fts\fi%
\immediate\write\ftfile{\noexpand\smallskip%
\noexpand\item{\noexpand\hyperdef\noexpand\hypernoname{footnote}
{\the\ftno}{f\the\ftno}:\ }\pctsign}\findarg}%
\def\footatend{\vfill\eject\immediate\closeout\ftfile{\parindent=20pt
\centerline{\bf Footnotes}\nobreak\bigskip\input \jobname.fts }}}
\def\footatend{}
%
%
\global\newcount\refno \global\refno=1
\newwrite\rfile
\def\ref{[\hyperref{}{reference}{\the\refno}{\the\refno}]\nref}
\def\nref#1{\DefWarn#1%
\xdef#1{[\noexpand\hyperref{}{reference}{\the\refno}{\the\refno}]}%
\writedef{#1\leftbracket#1}%
\ifnum\refno=1\immediate\openout\rfile=\jobname.refs\fi
\chardef\wfile=\rfile\immediate\write\rfile{\noexpand\item{[\noexpand\hyperdef%
\noexpand\hypernoname{reference}{\the\refno}{\the\refno}]\ }%
\reflabeL{#1\hskip.31in}\pctsign}\global\advance\refno by1\findarg}
\def\findarg#1#{\begingroup\obeylines\newlinechar=`\^^M\pass@rg}
{\obeylines\gdef\pass@rg#1{\writ@line\relax #1^^M\hbox{}^^M}%
\gdef\writ@line#1^^M{\expandafter\toks0\expandafter{\striprel@x #1}%
\edef\next{\the\toks0}\ifx\next\em@rk\let\next=\endgroup\else\ifx\next\empty%
\else\immediate\write\wfile{\the\toks0}\fi\let\next=\writ@line\fi\next\relax}}
\def\striprel@x#1{} \def\em@rk{\hbox{}}
\def\lref{\begingroup\obeylines\lr@f}
\def\lr@f#1#2{\DefWarn#1\gdef#1{\let#1=\UNd@FiNeD\ref#1{#2}}\endgroup\unskip}

\def\addref#1{\immediate\write\rfile{\noexpand\item{}#1}} 
\def\listrefs{\footatend\vfill\supereject\immediate\closeout\rfile\writestoppt
\baselineskip=\footskip\centerline{{\bf References}}\bigskip{\parindent=20pt%
\frenchspacing\escapechar=` \input \jobname.refs\vfill\eject}\nonfrenchspacing}
\def\startrefs#1{\immediate\openout\rfile=\jobname.refs\refno=#1}
\def\xref{\expandafter\xr@f}\def\xr@f[#1]{#1}
\def\refs#1{\count255=1[\r@fs #1{\hbox{}}]}
\def\r@fs#1{\ifx\UNd@FiNeD#1\message{reflabel \string#1 is undefined.}%
\nref#1{need to supply reference \string#1.}\fi%
\vphantom{\hphantom{#1}}{\let\hyperref=\relax\xdef\next{#1}}%
\ifx\next\em@rk\def\next{}%
\else\ifx\next#1\ifodd\count255\relax\xref#1\count255=0\fi%
\else#1\count255=1\fi\let\next=\r@fs\fi\next}
%

%
\newwrite\ffile\global\newcount\figno \global\figno=1
\def\fig{fig.~\hyperref{}{figure}{\the\figno}{\the\figno}\nfig}
\def\nfig#1{\DefWarn#1%
\xdef#1{fig.~\noexpand\hyperref{}{figure}{\the\figno}{\the\figno}}%
\writedef{#1\leftbracket fig.\noexpand~\xfig#1}%
\ifnum\figno=1\immediate\openout\ffile=\jobname.figs\fi\chardef\wfile=\ffile%
{\let\hyperref=\relax
\immediate\write\ffile{\noexpand\medskip\noexpand\item{Fig.\ %
\noexpand\hyperdef\noexpand\hypernoname{figure}{\the\figno}{\the\figno}. }
\reflabeL{#1\hskip.55in}\pctsign}}\global\advance\figno by1\findarg}
\def\listfigs{\vfill\eject\immediate\closeout\ffile{\parindent40pt
\baselineskip14pt\centerline{{\bf Figure Captions}}\nobreak\medskip
\escapechar=` \input \jobname.figs\vfill\eject}}
\def\xfig{\expandafter\xf@g}\def\xf@g fig.\penalty\@M\ {}
\def\figs#1{figs.~\f@gs #1{\hbox{}}}
\def\f@gs#1{{\let\hyperref=\relax\xdef\next{#1}}\ifx\next\em@rk\def\next{}\else
\ifx\next#1\xfig #1\else#1\fi\let\next=\f@gs\fi\next}
\def\figin{\epsfcheck\figin}\def\figins{\epsfcheck\figins}
\def\epsfcheck{\ifx\epsfbox\UNd@FiNeD
\message{(NO epsf.tex, FIGURES WILL BE IGNORED)}
\gdef\figin##1{\vskip2in}\gdef\figins##1{\hskip.5in}
\else\message{(FIGURES WILL BE INCLUDED)}%
\gdef\figin##1{##1}\gdef\figins##1{##1}\fi}
\def\DefWarn#1{}
\def\figinsert{\goodbreak\midinsert}
\def\ifig#1#2#3{\DefWarn#1\xdef#1{fig.~\noexpand\hyperref{}{figure}%
{\the\figno}{\the\figno}}\writedef{#1\leftbracket fig.\noexpand~\xfig#1}%
\figinsert\figin{\centerline{#3}}\medskip\centerline{\vbox{\baselineskip12pt
\advance\hsize by -1truein\noindent\wrlabeL{#1=#1}\footnotefont%
{\bf Fig.~\hyperdef\hypernoname{figure}{\the\figno}{\the\figno}:} #2}}
\bigskip\endinsert\global\advance\figno by1}
\newwrite\lfile
{\escapechar-1\xdef\pctsign{\string\%}\xdef\leftbracket{\string\{}
\xdef\rightbracket{\string\}}\xdef\numbersign{\string\#}}
\def\writedefs{\immediate\openout\lfile=\jobname.defs \def\writedef##1{%
{\let\hyperref=\relax\let\hyperdef=\relax\let\hypernoname=\relax
 \immediate\write\lfile{\string\def\string##1\rightbracket}}}}%
\def\writestop{\def\writestoppt{\immediate\write\lfile{\string\pageno
 \the\pageno\string\startrefs\leftbracket\the\refno\rightbracket
 \string\def\string\secsym\leftbracket\secsym\rightbracket
 \string\secno\the\secno\string\meqno\the\meqno}\immediate\closeout\lfile}}
\def\writestoppt{}\def\writedef#1{}
\def\seclab#1{\DefWarn#1%
\xdef #1{\noexpand\hyperref{}{section}{\the\secno}{\the\secno}}%
\writedef{#1\leftbracket#1}\wrlabeL{#1=#1}}
\def\subseclab#1{\DefWarn#1%
\xdef #1{\noexpand\hyperref{}{subsection}{\secn@m.\the\subsecno}%
{\secn@m.\the\subsecno}}\writedef{#1\leftbracket#1}\wrlabeL{#1=#1}}
\def\applab#1{\DefWarn#1%
\xdef #1{\noexpand\hyperref{}{appendix}{\secn@m}{\secn@m}}%
\writedef{#1\leftbracket#1}\wrlabeL{#1=#1}}
\newwrite\tfile \def\writetoca#1{}
\def\leaderfill{\leaders\hbox to 1em{\hss.\hss}\hfill}
\def\writetoc{\immediate\openout\tfile=\jobname.toc
   \def\writetoca##1{{\edef\next{\write\tfile{\noindent ##1
   \string\leaderfill {\string\hyperref{}{page}{\noexpand\number\pageno}%
                       {\noexpand\number\pageno}} \par}}\next}}}
\newread\ch@ckfile
\def\listtoc{\immediate\closeout\tfile\immediate\openin\ch@ckfile=\jobname.toc
\ifeof\ch@ckfile\message{no file \jobname.toc, no table of contents this pass}%
\else\closein\ch@ckfile\centerline{\bf Contents}\nobreak\medskip%
{\baselineskip=12pt\footnotefont\parskip=0pt\catcode`\@=11\input\jobname.toc
\catcode`\@=12\bigbreak\bigskip}\fi}
\catcode`\@=12 
%
\edef\tfontsize{\ifx\answ\bigans scaled\magstep3\else scaled\magstep4\fi}
\font\titlerm=cmr10 \tfontsize \font\titlerms=cmr7 \tfontsize
\font\titlermss=cmr5 \tfontsize \font\titlei=cmmi10 \tfontsize
\font\titleis=cmmi7 \tfontsize \font\titleiss=cmmi5 \tfontsize
\font\titlesy=cmsy10 \tfontsize \font\titlesys=cmsy7 \tfontsize
\font\titlesyss=cmsy5 \tfontsize \font\titleit=cmti10 \tfontsize
\skewchar\titlei='177 \skewchar\titleis='177 \skewchar\titleiss='177
\skewchar\titlesy='60 \skewchar\titlesys='60 \skewchar\titlesyss='60
\def\titlefont{\def\rm{\fam0\titlerm}
\textfont0=\titlerm \scriptfont0=\titlerms \scriptscriptfont0=\titlermss
\textfont1=\titlei \scriptfont1=\titleis \scriptscriptfont1=\titleiss
\textfont2=\titlesy \scriptfont2=\titlesys \scriptscriptfont2=\titlesyss
\textfont\itfam=\titleit \def\it{\fam\itfam\titleit}\rm}
 \ifx\answ\bigans\else scaled\magstep1\fi
\ifx\answ\bigans\def\abstractfont{\tenpoint}\else
\font\absit=cmti10 scaled \magstep1
\font\abssl=cmsl10 scaled \magstep1
\font\absrm=cmr10 scaled\magstep1 \font\absrms=cmr7 scaled\magstep1
\font\absrmss=cmr5 scaled\magstep1 \font\absi=cmmi10 scaled\magstep1
\font\absis=cmmi7 scaled\magstep1 \font\absiss=cmmi5 scaled\magstep1
\font\abssy=cmsy10 scaled\magstep1 \font\abssys=cmsy7 scaled\magstep1
\font\abssyss=cmsy5 scaled\magstep1 \font\absbf=cmbx10 scaled\magstep1
\skewchar\absi='177 \skewchar\absis='177 \skewchar\absiss='177
\skewchar\abssy='60 \skewchar\abssys='60 \skewchar\abssyss='60
\def\abstractfont{\def\rm{\fam0\absrm}
\textfont0=\absrm \scriptfont0=\absrms \scriptscriptfont0=\absrmss
\textfont1=\absi \scriptfont1=\absis \scriptscriptfont1=\absiss
\textfont2=\abssy \scriptfont2=\abssys \scriptscriptfont2=\abssyss
\textfont\itfam=\absit \def\it{\fam\itfam\absit}\def\footnotefont{\tenpoint}%
\textfont\slfam=\abssl \def\sl{\fam\slfam\abssl}%
\textfont\bffam=\absbf \def\bf{\fam\bffam\absbf}\rm}\fi
\def\tenpoint{\def\rm{\fam0\tenrm}
\textfont0=\tenrm \scriptfont0=\sevenrm \scriptscriptfont0=\fiverm
\textfont1=\teni  \scriptfont1=\seveni  \scriptscriptfont1=\fivei
\textfont2=\tensy \scriptfont2=\sevensy \scriptscriptfont2=\fivesy
\textfont\itfam=\tenit \def\it{\fam\itfam\tenit}\def\footnotefont{\ninepoint}%
\textfont\bffam=\tenbf \def\bf{\fam\bffam\tenbf}\def\sl{\fam\slfam\tensl}\rm}
\font\ninerm=cmr9 \font\sixrm=cmr6 \font\ninei=cmmi9 \font\sixi=cmmi6
\font\ninesy=cmsy9 \font\sixsy=cmsy6 \font\ninebf=cmbx9
\font\nineit=cmti9 \font\ninesl=cmsl9 \skewchar\ninei='177
\skewchar\sixi='177 \skewchar\ninesy='60 \skewchar\sixsy='60
\def\ninepoint{\def\rm{\fam0\ninerm}
\textfont0=\ninerm \scriptfont0=\sixrm \scriptscriptfont0=\fiverm
\textfont1=\ninei \scriptfont1=\sixi \scriptscriptfont1=\fivei
\textfont2=\ninesy \scriptfont2=\sixsy \scriptscriptfont2=\fivesy
\textfont\itfam=\ninei \def\it{\fam\itfam\nineit}\def\sl{\fam\slfam\ninesl}%
\textfont\bffam=\ninebf \def\bf{\fam\bffam\ninebf}\rm}
%
%

\hyphenation{anom-aly anom-alies coun-ter-term coun-ter-terms}
\def\inv{^{\raise.15ex\hbox{${\scriptscriptstyle -}$}\kern-.05em 1}}

\def\Dsl{\,\raise.15ex\hbox{/}\mkern-13.5mu D} 
\def\dsl{\raise.15ex\hbox{/}\kern-.57em\partial}

\def\lspace{\ifx\answ\bigans{}\else\qquad\fi}
\def\lbspace{\ifx\answ\bigans{}\else\hskip-.2in\fi} 
\def\boxeqn#1{\vcenter{\vbox{\hrule\hbox{\vrule\kern3pt\vbox{\kern3pt
	\hbox{${\displaystyle #1}$}\kern3pt}\kern3pt\vrule}\hrule}}}
\def\mbox#1#2{\vcenter{\hrule \hbox{\vrule height#2in
		\kern#1in \vrule} \hrule}}  
%

\def\darr#1{\raise1.5ex\hbox{$\leftrightarrow$}\mkern-16.5mu #1}

\def\roughly#1{\raise.3ex\hbox{$#1$\kern-.75em\lower1ex\hbox{$\sim$}}}

\def\smallfig#1#2#3{\DefWarn#1\xdef#1{fig.~\the\figno}
\writedef{#1\leftbracket fig.\noexpand~\the\figno}%
\figinsert\figin{\centerline{#3}}\medskip\centerline{\vbox{
\baselineskip12pt\advance\hsize by -1truein
\noindent\footnotefont{\bf Fig.~\the\figno:} #2}}
\endinsert\global\advance\figno by1}

\def\bb{
\font\tenmsb=msbm10
\font\sevenmsb=msbm7
\font\fivemsb=msbm5
\textfont1=\tenmsb
\scriptfont1=\sevenmsb
\scriptscriptfont1=\fivemsb
}

\input amssym

%
%
\ifx\pdfoutput\undefined
\input epsf
\def\fig#1{\epsfbox{#1.eps}}
\def\figscale#1#2{\epsfxsize=#2\epsfbox{#1.eps}}
%
%
\else
\def\fig#1{\pdfximage {#1.pdf}\pdfrefximage\pdflastximage}
\def\figscale#1#2{\pdfximage width#2 {#1.pdf}\pdfrefximage\pdflastximage}
\fi

\def\IZ{\relax\ifmmode\mathchoice
{\hbox{\cmss Z\kern-.4em Z}}{\hbox{\cmss Z\kern-.4em Z}} {\lower.9pt\hbox{\cmsss Z\kern-.4em Z}}
{\lower1.2pt\hbox{\cmsss Z\kern-.4em Z}}\else{\cmss Z\kern-.4em Z}\fi}

\newif\ifdraft\draftfalse
\newif\ifinter\interfalse
\ifdraft\draftmode\else\interfalse\fi
\def\journal#1&#2(#3){\unskip, \sl #1\ \bf #2 \rm(19#3) }
\def\andjournal#1&#2(#3){\sl #1~\bf #2 \rm (19#3) }

\def\frac#1#2{{#1\over#2}}

\def\inbar{\,\vrule height1.5ex width.4pt depth0pt}
\def\IC{\relax\hbox{$\inbar\kern-.3em{\rm C}$}}
\def\IR{\relax{\rm I\kern-.18em R}}
\def\IP{\relax{\rm I\kern-.18em P}}
\def\Z{{\bf Z}}

%
%


%
\catcode`\@=11
\def\slash#1{\mathord{\mathpalette\c@ncel{#1}}}
\overfullrule=0pt

\def\underrel#1\over#2{\mathrel{\mathop{\kern\z@#1}\limits_{#2}}}

\catcode`\@=12


%


\def\[{[}
\def\]{]}

\def\comment#1{ }

%
\def\draftnote#1{\ifdraft{\baselineskip2ex
                 \vbox{\kern1em\hrule\hbox{\vrule\kern1em\vbox{\kern1ex
                 \noindent \underbar{NOTE}: #1
             \vskip1ex}\kern1em\vrule}\hrule}}\fi}
\def\internote#1{\ifinter{\baselineskip2ex
                 \vbox{\kern1em\hrule\hbox{\vrule\kern1em\vbox{\kern1ex
                 \noindent \underbar{Internal Note}: #1
             \vskip1ex}\kern1em\vrule}\hrule}}\fi}

%
%



%
%
%
%

%

\def\inv{^{-1}}


\def\1{{\ds 1}}

\def\C{\hbox{$\bb C$}}

\def\Z{\hbox{$\bb Z$}}

\def\P{\hbox{$\bb P$}}

\def\S{\hbox{$\bb S$}}

\newfam\frakfam
\font\teneufm=eufm10
\font\seveneufm=eufm7
\font\fiveeufm=eufm5
\textfont\frakfam=\teneufm
\scriptfont\frakfam=\seveneufm
\scriptscriptfont\frakfam=\fiveeufm

\lref\NiarchosAH{
  V.~Niarchos,
  ``Seiberg dualities and the 3d/4d connection,''
JHEP {\bf 1207}, 075 (2012).
[arXiv:1205.2086 [hep-th]].
}

\lref\AharonyGP{
  O.~Aharony,
  ``IR duality in d = 3 N=2 supersymmetric USp(2N(c)) and U(N(c)) gauge theories,''
Phys.\ Lett.\ B {\bf 404}, 71 (1997).
[hep-th/9703215].
}

\lref\AffleckAS{
  I.~Affleck, J.~A.~Harvey and E.~Witten,
  ``Instantons and (Super)Symmetry Breaking in (2+1)-Dimensions,''
Nucl.\ Phys.\ B {\bf 206}, 413 (1982)..
}

\lref\IntriligatorID{
  K.~A.~Intriligator and N.~Seiberg,
  ``Duality, monopoles, dyons, confinement and oblique confinement in supersymmetric SO(N(c)) gauge theories,''
Nucl.\ Phys.\ B {\bf 444}, 125 (1995).
[hep-th/9503179].
}

\lref\PasquettiFJ{
  S.~Pasquetti,
  ``Factorisation of N = 2 Theories on the Squashed 3-Sphere,''
JHEP {\bf 1204}, 120 (2012).
[arXiv:1111.6905 [hep-th]].
}

\lref\BeemMB{
  C.~Beem, T.~Dimofte and S.~Pasquetti,
  ``Holomorphic Blocks in Three Dimensions,''
[arXiv:1211.1986 [hep-th]].
}

\lref\SeibergPQ{
  N.~Seiberg,
  ``Electric - magnetic duality in supersymmetric nonAbelian gauge theories,''
Nucl.\ Phys.\ B {\bf 435}, 129 (1995).
[hep-th/9411149].
}

\lref\AharonyBX{
  O.~Aharony, A.~Hanany, K.~A.~Intriligator, N.~Seiberg and M.~J.~Strassler,
  ``Aspects of N=2 supersymmetric gauge theories in three-dimensions,''
Nucl.\ Phys.\ B {\bf 499}, 67 (1997).
[hep-th/9703110].
}

\lref\IntriligatorNE{
  K.~A.~Intriligator and P.~Pouliot,
  ``Exact superpotentials, quantum vacua and duality in supersymmetric SP(N(c)) gauge theories,''
Phys.\ Lett.\ B {\bf 353}, 471 (1995).
[hep-th/9505006].
}

\lref\KarchUX{
  A.~Karch,
  ``Seiberg duality in three-dimensions,''
Phys.\ Lett.\ B {\bf 405}, 79 (1997).
[hep-th/9703172].
}

\lref\SafdiRE{
  B.~R.~Safdi, I.~R.~Klebanov and J.~Lee,
  ``A Crack in the Conformal Window,''
[arXiv:1212.4502 [hep-th]].
}

\lref\SchweigertTG{
  C.~Schweigert,
  ``On moduli spaces of flat connections with nonsimply connected structure group,''
Nucl.\ Phys.\ B {\bf 492}, 743 (1997).
[hep-th/9611092].
}

\lref\GiveonZN{
  A.~Giveon and D.~Kutasov,
  ``Seiberg Duality in Chern-Simons Theory,''
Nucl.\ Phys.\ B {\bf 812}, 1 (2009).
[arXiv:0808.0360 [hep-th]].
}

\lref\Spiridonov{
  Spiridonov, V.~P.,
  ``Aspects of elliptic hypergeometric functions,''
[arXiv:1307.2876 [math.CA]].
}

\lref\GaiottoBE{
  D.~Gaiotto, G.~W.~Moore and A.~Neitzke,
  ``Framed BPS States,''
[arXiv:1006.0146 [hep-th]].
}

\lref\AldayRS{
  L.~F.~Alday, M.~Bullimore and M.~Fluder,
  ``On S-duality of the Superconformal Index on Lens Spaces and 2d TQFT,''
JHEP {\bf 1305}, 122 (2013).
[arXiv:1301.7486 [hep-th]].
}

\lref\RazamatJXA{
  S.~S.~Razamat and M.~Yamazaki,
  ``S-duality and the N=2 Lens Space Index,''
[arXiv:1306.1543 [hep-th]].
}

\lref\NiarchosAH{
  V.~Niarchos,
  ``Seiberg dualities and the 3d/4d connection,''
JHEP {\bf 1207}, 075 (2012).
[arXiv:1205.2086 [hep-th]].
}

\lref\almost{
  A.~Borel, R.~Friedman, J.~W.~Morgan,
  ``Almost commuting elements in compact Lie groups,''
arXiv:math/9907007.
}

\lref\KapustinJM{
  A.~Kapustin and B.~Willett,
  ``Generalized Superconformal Index for Three Dimensional Field Theories,''
[arXiv:1106.2484 [hep-th]].
}

\lref\AharonyGP{
  O.~Aharony,
  ``IR duality in d = 3 N=2 supersymmetric USp(2N(c)) and U(N(c)) gauge theories,''
Phys.\ Lett.\ B {\bf 404}, 71 (1997).
[hep-th/9703215].
}

\lref\FestucciaWS{
  G.~Festuccia and N.~Seiberg,
  ``Rigid Supersymmetric Theories in Curved Superspace,''
JHEP {\bf 1106}, 114 (2011).
[arXiv:1105.0689 [hep-th]].
}

\lref\RomelsbergerEG{
  C.~Romelsberger,
  ``Counting chiral primaries in N = 1, d=4 superconformal field theories,''
Nucl.\ Phys.\ B {\bf 747}, 329 (2006).
[hep-th/0510060].
}

\lref\KapustinKZ{
  A.~Kapustin, B.~Willett and I.~Yaakov,
  ``Exact Results for Wilson Loops in Superconformal Chern-Simons Theories with Matter,''
JHEP {\bf 1003}, 089 (2010).
[arXiv:0909.4559 [hep-th]].
}

\lref\DolanQI{
  F.~A.~Dolan and H.~Osborn,
  ``Applications of the Superconformal Index for Protected Operators and q-Hypergeometric Identities to N=1 Dual Theories,''
Nucl.\ Phys.\ B {\bf 818}, 137 (2009).
[arXiv:0801.4947 [hep-th]].
}

\lref\GaddeIA{
  A.~Gadde and W.~Yan,
  ``Reducing the 4d Index to the $S^3$ Partition Function,''
JHEP {\bf 1212}, 003 (2012).
[arXiv:1104.2592 [hep-th]].
}

\lref\DolanRP{
  F.~A.~H.~Dolan, V.~P.~Spiridonov and G.~S.~Vartanov,
  ``From 4d superconformal indices to 3d partition functions,''
Phys.\ Lett.\ B {\bf 704}, 234 (2011).
[arXiv:1104.1787 [hep-th]].
}

\lref\ImamuraUW{
  Y.~Imamura,
 ``Relation between the 4d superconformal index and the $S^3$ partition function,''
JHEP {\bf 1109}, 133 (2011).
[arXiv:1104.4482 [hep-th]].
}

\lref\BeemYN{
  C.~Beem and A.~Gadde,
  ``The $N=1$ superconformal index for class $S$ fixed points,''
JHEP {\bf 1404}, 036 (2014).
[arXiv:1212.1467 [hep-th]].
}

\lref\HamaEA{
  N.~Hama, K.~Hosomichi and S.~Lee,
  ``SUSY Gauge Theories on Squashed Three-Spheres,''
JHEP {\bf 1105}, 014 (2011).
[arXiv:1102.4716 [hep-th]].
}

\lref\GaddeEN{
  A.~Gadde, L.~Rastelli, S.~S.~Razamat and W.~Yan,
  ``On the Superconformal Index of N=1 IR Fixed Points: A Holographic Check,''
JHEP {\bf 1103}, 041 (2011).
[arXiv:1011.5278 [hep-th]].
}

\lref\EagerHX{
  R.~Eager, J.~Schmude and Y.~Tachikawa,
  ``Superconformal Indices, Sasaki-Einstein Manifolds, and Cyclic Homologies,''
[arXiv:1207.0573 [hep-th]].
}

\lref\AffleckAS{
  I.~Affleck, J.~A.~Harvey and E.~Witten,
  ``Instantons and (Super)Symmetry Breaking in (2+1)-Dimensions,''
Nucl.\ Phys.\ B {\bf 206}, 413 (1982)..
}

\lref\SeibergPQ{
  N.~Seiberg,
  ``Electric - magnetic duality in supersymmetric nonAbelian gauge theories,''
Nucl.\ Phys.\ B {\bf 435}, 129 (1995).
[hep-th/9411149].
}

\lref\BahDG{
  I.~Bah, C.~Beem, N.~Bobev and B.~Wecht,
  ``Four-Dimensional SCFTs from M5-Branes,''
JHEP {\bf 1206}, 005 (2012).
[arXiv:1203.0303 [hep-th]].
}

\lref\debult{
  F.~van~de~Bult,
  ``Hyperbolic Hypergeometric Functions,''
University of Amsterdam Ph.D. thesis
}

\lref\Shamirthesis{
  I.~Shamir,
  ``Aspects of three dimensional Seiberg duality,''
  M. Sc. thesis submitted to the Weizmann Institute of Science, April 2010.
  }

\lref\slthreeZ{
  J.~Felder, A.~Varchenko,
  ``The elliptic gamma function and $SL(3,Z) \times Z^3$,'' $\;\;$
[arXiv:math/0001184].
}

\lref\BeniniNC{
  F.~Benini, T.~Nishioka and M.~Yamazaki,
  ``4d Index to 3d Index and 2d TQFT,''
Phys.\ Rev.\ D {\bf 86}, 065015 (2012).
[arXiv:1109.0283 [hep-th]].
}

\lref\GaiottoWE{
  D.~Gaiotto,
  ``N=2 dualities,''
  JHEP {\bf 1208}, 034 (2012).
  [arXiv:0904.2715 [hep-th]].
}

\lref\SpiridonovZA{
  V.~P.~Spiridonov and G.~S.~Vartanov,
  ``Elliptic Hypergeometry of Supersymmetric Dualities,''
Commun.\ Math.\ Phys.\  {\bf 304}, 797 (2011).
[arXiv:0910.5944 [hep-th]].
}

\lref\BeniniMF{
  F.~Benini, C.~Closset and S.~Cremonesi,
  ``Comments on 3d Seiberg-like dualities,''
JHEP {\bf 1110}, 075 (2011).
[arXiv:1108.5373 [hep-th]].
}

\lref\ClossetVP{
  C.~Closset, T.~T.~Dumitrescu, G.~Festuccia, Z.~Komargodski and N.~Seiberg,
  ``Comments on Chern-Simons Contact Terms in Three Dimensions,''
JHEP {\bf 1209}, 091 (2012).
[arXiv:1206.5218 [hep-th]].
}

\lref\SpiridonovHF{
  V.~P.~Spiridonov and G.~S.~Vartanov,
  ``Elliptic hypergeometry of supersymmetric dualities II. Orthogonal groups, knots, and vortices,''
[arXiv:1107.5788 [hep-th]].
}

\lref\SpiridonovWW{
  V.~P.~Spiridonov and G.~S.~Vartanov,
  ``Elliptic hypergeometric integrals and 't Hooft anomaly matching conditions,''
JHEP {\bf 1206}, 016 (2012).
[arXiv:1203.5677 [hep-th]].
}

\lref\DimoftePY{
  T.~Dimofte, D.~Gaiotto and S.~Gukov,
  ``3-Manifolds and 3d Indices,''
[arXiv:1112.5179 [hep-th]].
}

\lref\KimWB{
  S.~Kim,
  ``The Complete superconformal index for N=6 Chern-Simons theory,''
Nucl.\ Phys.\ B {\bf 821}, 241 (2009), [Erratum-ibid.\ B {\bf 864}, 884 (2012)].
[arXiv:0903.4172 [hep-th]].
}

\lref\WillettGP{
  B.~Willett and I.~Yaakov,
  ``N=2 Dualities and Z Extremization in Three Dimensions,''
[arXiv:1104.0487 [hep-th]].
}

\lref\ImamuraSU{
  Y.~Imamura and S.~Yokoyama,
  ``Index for three dimensional superconformal field theories with general R-charge assignments,''
JHEP {\bf 1104}, 007 (2011).
[arXiv:1101.0557 [hep-th]].
}

\lref\FreedYA{
  D.~S.~Freed, G.~W.~Moore and G.~Segal,
  ``The Uncertainty of Fluxes,''
Commun.\ Math.\ Phys.\  {\bf 271}, 247 (2007).
[hep-th/0605198].
}

\lref\HwangQT{
  C.~Hwang, H.~Kim, K.~-J.~Park and J.~Park,
  ``Index computation for 3d Chern-Simons matter theory: test of Seiberg-like duality,''
JHEP {\bf 1109}, 037 (2011).
[arXiv:1107.4942 [hep-th]].
}

\lref\GreenDA{
  D.~Green, Z.~Komargodski, N.~Seiberg, Y.~Tachikawa and B.~Wecht,
  ``Exactly Marginal Deformations and Global Symmetries,''
JHEP {\bf 1006}, 106 (2010).
[arXiv:1005.3546 [hep-th]].
}

\lref\GaiottoXA{
  D.~Gaiotto, L.~Rastelli and S.~S.~Razamat,
  ``Bootstrapping the superconformal index with surface defects,''
[arXiv:1207.3577 [hep-th]].
}

\lref\IntriligatorID{
  K.~A.~Intriligator and N.~Seiberg,
  ``Duality, monopoles, dyons, confinement and oblique confinement in supersymmetric SO(N(c)) gauge theories,''
Nucl.\ Phys.\ B {\bf 444}, 125 (1995).
[hep-th/9503179].
}

\lref\SeibergNZ{
  N.~Seiberg and E.~Witten,
  ``Gauge dynamics and compactification to three-dimensions,''
In *Saclay 1996, The mathematical beauty of physics* 333-366.
[hep-th/9607163].
}

\lref\KinneyEJ{
  J.~Kinney, J.~M.~Maldacena, S.~Minwalla and S.~Raju,
  ``An Index for 4 dimensional super conformal theories,''
  Commun.\ Math.\ Phys.\  {\bf 275}, 209 (2007).
  [hep-th/0510251].
}

\lref\NakayamaUR{
  Y.~Nakayama,
  ``Index for supergravity on AdS(5) x T**1,1 and conifold gauge theory,''
Nucl.\ Phys.\ B {\bf 755}, 295 (2006).
[hep-th/0602284].
}

\lref\GaddeKB{
  A.~Gadde, E.~Pomoni, L.~Rastelli and S.~S.~Razamat,
  ``S-duality and 2d Topological QFT,''
JHEP {\bf 1003}, 032 (2010).
[arXiv:0910.2225 [hep-th]].
}

\lref\GaddeTE{
  A.~Gadde, L.~Rastelli, S.~S.~Razamat and W.~Yan,
  ``The Superconformal Index of the $E_6$ SCFT,''
JHEP {\bf 1008}, 107 (2010).
[arXiv:1003.4244 [hep-th]].
}

\lref\AharonyCI{
  O.~Aharony and I.~Shamir,
  ``On $O(N_c)$ d=3 N=2 supersymmetric QCD Theories,''
JHEP {\bf 1112}, 043 (2011).
[arXiv:1109.5081 [hep-th]].
}

\lref\GiveonSR{
  A.~Giveon and D.~Kutasov,
  ``Brane dynamics and gauge theory,''
Rev.\ Mod.\ Phys.\  {\bf 71}, 983 (1999).
[hep-th/9802067].
}

\lref\SpiridonovQV{
  V.~P.~Spiridonov and G.~S.~Vartanov,
  ``Superconformal indices of ${\cal N}=4$ SYM field theories,''
Lett.\ Math.\ Phys.\  {\bf 100}, 97 (2012).
[arXiv:1005.4196 [hep-th]].
}
\lref\GaddeUV{
  A.~Gadde, L.~Rastelli, S.~S.~Razamat and W.~Yan,
  ``Gauge Theories and Macdonald Polynomials,''
Commun.\ Math.\ Phys.\  {\bf 319}, 147 (2013).
[arXiv:1110.3740 [hep-th]].
}
\lref\KapustinGH{
  A.~Kapustin,
  ``Seiberg-like duality in three dimensions for orthogonal gauge groups,''
[arXiv:1104.0466 [hep-th]].
}

\lref\orthogpaper{O. Aharony, S. S. Razamat, N.~Seiberg and B.~Willett, 
``3d dualities from 4d dualities for orthogonal groups,''
[arXiv:1307.0511 [hep-th]].
}

\lref\readinglines{
  O.~Aharony, N.~Seiberg and Y.~Tachikawa,
  ``Reading between the lines of four-dimensional gauge theories,''
[arXiv:1305.0318 [hep-th]].
}

\lref\WittenNV{
  E.~Witten,
  ``Supersymmetric index in four-dimensional gauge theories,''
Adv.\ Theor.\ Math.\ Phys.\  {\bf 5}, 841 (2002).
[hep-th/0006010].
}

\lref\GaddeUV{
  A.~Gadde, L.~Rastelli, S.~S.~Razamat and W.~Yan,
  ``Gauge Theories and Macdonald Polynomials,''
Commun.\ Math.\ Phys.\  {\bf 319}, 147 (2013).
[arXiv:1110.3740 [hep-th]].
}

\lref\GaddeIK{
  A.~Gadde, L.~Rastelli, S.~S.~Razamat and W.~Yan,
  ``The 4d Superconformal Index from q-deformed 2d Yang-Mills,''
Phys.\ Rev.\ Lett.\  {\bf 106}, 241602 (2011).
[arXiv:1104.3850 [hep-th]].
}

\lref\GaiottoXA{
  D.~Gaiotto, L.~Rastelli and S.~S.~Razamat,
  ``Bootstrapping the superconformal index with surface defects,''
JHEP {\bf 1301}, 022 (2013).
[arXiv:1207.3577 [hep-th]].
}

\lref\GaiottoUQ{
  D.~Gaiotto and S.~S.~Razamat,
  ``Exceptional Indices,''
JHEP {\bf 1205}, 145 (2012).
[arXiv:1203.5517 [hep-th]].
}

\lref\RazamatUV{
  S.~S.~Razamat,
  ``On a modular property of N=2 superconformal theories in four dimensions,''
JHEP {\bf 1210}, 191 (2012).
[arXiv:1208.5056 [hep-th]].
}

\lref\noumi{
  Y.~Komori, M.~Noumi, J.~Shiraishi,
  ``Kernel Functions for Difference Operators of Ruijsenaars Type and Their Applications,''
SIGMA 5 (2009), 054.
[arXiv:0812.0279 [math.QA]].
}

\lref\SpirWarnaar{
  V.~P.~Spiridonov and S.~O.~Warnaar,
  ``Inversions of integral operators and elliptic beta integrals on root systems,''
Adv. Math. 207 (2006), 91-132
[arXiv:math/0411044].
}

\lref\RazamatJXA{
  S.~S.~Razamat and M.~Yamazaki,
  ``S-duality and the N=2 Lens Space Index,''
[arXiv:1306.1543 [hep-th]].
}

\lref\RazamatOPA{
  S.~S.~Razamat and B.~Willett,
  ``Global Properties of Supersymmetric Theories and the Lens Space,''
[arXiv:1307.4381 [hep-th]].
}

\lref\GaddeTE{
  A.~Gadde, L.~Rastelli, S.~S.~Razamat and W.~Yan,
  ``The Superconformal Index of the $E_6$ SCFT,''
JHEP {\bf 1008}, 107 (2010).
[arXiv:1003.4244 [hep-th]].
}

\lref\deBult{
  F.~J.~van~de~Bult,
  ``An elliptic hypergeometric integral with $W(F_4)$ symmetry,''
The Ramanujan Journal, Volume 25, Issue 1 (2011)
[arXiv:0909.4793[math.CA]].
}

\lref\GaddeKB{
  A.~Gadde, E.~Pomoni, L.~Rastelli and S.~S.~Razamat,
  ``S-duality and 2d Topological QFT,''
JHEP {\bf 1003}, 032 (2010).
[arXiv:0910.2225 [hep-th]].
}

\lref\ArgyresCN{
  P.~C.~Argyres and N.~Seiberg,
  ``S-duality in N=2 supersymmetric gauge theories,''
JHEP {\bf 0712}, 088 (2007).
[arXiv:0711.0054 [hep-th]].
}

\lref\SpirWarnaar{
  V.~P.~Spiridonov and S.~O.~Warnaar,
  ``Inversions of integral operators and elliptic beta integrals on root systems,''
Adv. Math. 207 (2006), 91-132
[arXiv:math/0411044].
}

\lref\GaiottoHG{
  D.~Gaiotto, G.~W.~Moore and A.~Neitzke,
  ``Wall-crossing, Hitchin Systems, and the WKB Approximation,''
[arXiv:0907.3987 [hep-th]].
}

\lref\RuijsenaarsVQ{
  S.~N.~M.~Ruijsenaars and H.~Schneider,
  ``A New Class Of Integrable Systems And Its Relation To Solitons,''
Annals Phys.\  {\bf 170}, 370 (1986).
}

\lref\RuijsenaarsPP{
  S.~N.~M.~Ruijsenaars,
  ``Complete Integrability Of Relativistic Calogero-moser Systems And Elliptic Function Identities,''
Commun.\ Math.\ Phys.\  {\bf 110}, 191 (1987).
}

\lref\HallnasNB{
  M.~Hallnas and S.~Ruijsenaars,
  ``Kernel functions and Baecklund transformations for relativistic Calogero-Moser and Toda systems,''
J.\ Math.\ Phys.\  {\bf 53}, 123512 (2012).
}

\lref\kernelA{
S.~Ruijsenaars,
  ``Elliptic integrable systems of Calogero-Moser type: Some new results on joint eigenfunctions'', in Proceedings of the 2004 Kyoto Workshop on "Elliptic integrable systems", (M. Noumi, K. Takasaki, Eds.), Rokko Lectures in Math., no. 18, Dept. of Math., Kobe Univ.
}

\lref\ellRSreview{
Y.~Komori and S.~Ruijsenaars,
  ``Elliptic integrable systems of Calogero-Moser type: A survey'', in Proceedings of the 2004 Kyoto Workshop on "Elliptic integrable systems", (M. Noumi, K. Takasaki, Eds.), Rokko Lectures in Math., no. 18, Dept. of Math., Kobe Univ.
}

\lref\langmann{
E.~Langmann,
  ``An explicit solution of the (quantum) elliptic Calogero-Sutherland model'', [arXiv:math-ph/0407050].
}

\lref\TachikawaWI{
  Y.~Tachikawa,
  ``4d partition function on $S^1 \times S^3$ and 2d Yang-Mills with nonzero area,''
PTEP {\bf 2013}, 013B01 (2013).
[arXiv:1207.3497 [hep-th]].
}

\lref\MinahanFG{
  J.~A.~Minahan and D.~Nemeschansky,
  ``An N=2 superconformal fixed point with E(6) global symmetry,''
Nucl.\ Phys.\ B {\bf 482}, 142 (1996).
[hep-th/9608047].
}

\lref\AldayKDA{
  L.~F.~Alday, M.~Bullimore, M.~Fluder and L.~Hollands,
  ``Surface defects, the superconformal index and q-deformed Yang-Mills,''
[arXiv:1303.4460 [hep-th]].
}

\lref\IntriligatorEX{
  K.~A.~Intriligator and N.~Seiberg,
Phys.\ Lett.\ B {\bf 387}, 513 (1996).
[hep-th/9607207].
}

\lref\FukudaJR{
  Y.~Fukuda, T.~Kawano and N.~Matsumiya,
  ``5D SYM and 2D q-Deformed YM,''
Nucl.\ Phys.\ B {\bf 869}, 493 (2013).
[arXiv:1210.2855 [hep-th]].
}

\lref\XieHS{
  D.~Xie,
  ``General Argyres-Douglas Theory,''
JHEP {\bf 1301}, 100 (2013).
[arXiv:1204.2270 [hep-th]].
}

\lref\DrukkerSR{
  N.~Drukker, T.~Okuda and F.~Passerini,
  ``Exact results for vortex loop operators in 3d supersymmetric theories,''
[arXiv:1211.3409 [hep-th]].
}

\lref\qinteg{
  M.~Rahman, A.~Verma,
  ``A q-integral representation of Rogers' q-ultraspherical polynomials and some applications,''
Constructive Approximation
1986, Volume 2, Issue 1.
}

\lref\qintegOK{
  A.~Okounkov,
  ``(Shifted) Macdonald Polynomials: q-Integral Representation and Combinatorial Formula,''
Compositio Mathematica
June 1998, Volume 112, Issue 2. 
[arXiv:q-alg/9605013].
}

\lref\macNest{
 H.~Awata, S.~Odake, J.~Shiraishi,
  ``Integral Representations of the Macdonald Symmetric Functions,''
Commun. Math. Phys. 179 (1996) 647.
[arXiv:q-alg/9506006].
}

\lref\BeemYN{
  C.~Beem and A.~Gadde,
JHEP {\bf 1404}, 036 (2014).
[arXiv:1212.1467 [hep-th]].
}

\lref\deBult{
  F.~J.~van~de~Bult,
  ``An elliptic hypergeometric integral with $W(F_4)$ symmetry,''
The Ramanujan Journal, Volume 25, Issue 1 (2011)
[arXiv:0909.4793[math.CA]].
}

\lref\Rains{
E.~M.~Rains,
  ``Transformations of elliptic hypergometric integrals,''
Annals of Mathematics, Volume  171, Issue 1 (2010)
[arXiv:math/0309252].
}

\lref\DimoftePD{
  T.~Dimofte and D.~Gaiotto,
  ``An E7 Surprise,''
JHEP {\bf 1210}, 129 (2012).
[arXiv:1209.1404 [hep-th]].
}

\lref\GaddeFMA{
  A.~Gadde, K.~Maruyoshi, Y.~Tachikawa and W.~Yan,
  ``New N=1 Dualities,''
JHEP {\bf 1306}, 056 (2013).
[arXiv:1303.0836 [hep-th]].
}

\lref\GorskyTN{
  A.~Gorsky,
  ``Dualities in integrable systems and N=2 SUSY theories,''
J.\ Phys.\ A {\bf 34}, 2389 (2001).
[hep-th/9911037].
}

\lref\FockAE{
  V.~Fock, A.~Gorsky, N.~Nekrasov and V.~Rubtsov,
  ``Duality in integrable systems and gauge theories,''
JHEP {\bf 0007}, 028 (2000).
[hep-th/9906235].
}

\lref\RazamatPTA{
  S.~S.~Razamat and B.~Willett,
  ``Down the rabbit hole with theories of class $ \cal S $,''
JHEP {\bf 1410}, 99 (2014).
[arXiv:1403.6107 [hep-th]].
}

\lref\RazamatL{ A.~Gadde, S.~S.~Razamat, and  B.~Willett,
  ``A ``Lagrangian'' for a non-Lagrangian theory,''
  Phys. Rev. Lett. {\bf 115}, 171604 (2015).
  [arXiv:1505.05834 [hep-th]].
  }
  
\lref\PestunZXK{
V.~Pestun  {\it et al.},
 ``Localization techniques in quantum field theories,''
[arXiv:1608.02952 [hep-th]].
}
  
  \lref\OhmoriPUA{
  K.~Ohmori, H.~Shimizu, Y.~Tachikawa and K.~Yonekura,
  ``6d ${\cal N}=(1,0)$ theories on $T^2$ and class S theories: Part I,''
JHEP {\bf 1507}, 014 (2015).
[arXiv:1503.06217 [hep-th]].
}

\lref\SeibergBD{
  N.~Seiberg,
 ``Five-dimensional SUSY field theories, nontrivial fixed points and string dynamics,''
Phys.\ Lett.\ B {\bf 388}, 753 (1996).
[hep-th/9608111].
}

\lref\OhmoriPIA{
  K.~Ohmori, H.~Shimizu, Y.~Tachikawa and K.~Yonekura,
  ``6d ${\cal N}=(1,\;0) $ theories on S$^{1}$ /T$^{2}$ and class S theories: part II,''
JHEP {\bf 1512}, 131 (2015).
[arXiv:1508.00915 [hep-th]].
}

\lref\DelZottoRCA{
  M.~Del Zotto, C.~Vafa and D.~Xie,
  ``Geometric engineering, mirror symmetry and $ 6{{d}}_{\left(1,0\right)}\to 4{{d}}_{\left({\cal N}=2\right)} $,''
JHEP {\bf 1511}, 123 (2015).
[arXiv:1504.08348 [hep-th]].
}

\lref\HananyPFA{
  A.~Hanany and K.~Maruyoshi,
  ``Chiral theories of class $ {\cal S} $,''
JHEP {\bf 1512}, 080 (2015).
[arXiv:1505.05053 [hep-th]].
}

\lref\ComanBQQ{
  I.~Coman, E.~Pomoni, M.~Taki and F.~Yagi,
  ``Spectral curves of ${\cal N}=1$ theories of class ${\cal S}_k$,''
[arXiv:1512.06079 [hep-th]].
}

\lref\FrancoJNA{
  S.~Franco, H.~Hayashi and A.~Uranga,
  ``Charting Class ${\cal S}_k$ Territory,''
Phys.\ Rev.\ D {\bf 92}, no. 4, 045004 (2015).
[arXiv:1504.05988 [hep-th]].
}

\lref\SpiridonovZR{
  V.~P.~Spiridonov and G.~S.~Vartanov,
 ``Superconformal indices for N = 1 theories with multiple duals,''
Nucl.\ Phys.\ B {\bf 824}, 192 (2010).
[arXiv:0811.1909 [hep-th]].
}

\lref\AharonyDHA{
  O.~Aharony, S.~S.~Razamat, N.~Seiberg and B.~Willett,
  ``3d dualities from 4d dualities,''
JHEP {\bf 1307}, 149 (2013).
[arXiv:1305.3924 [hep-th]].}

\lref\CsakiCU{
  C.~Csaki, M.~Schmaltz, W.~Skiba and J.~Terning,
  ``Selfdual N=1 SUSY gauge theories,''
Phys.\ Rev.\ D {\bf 56}, 1228 (1997).
[hep-th/9701191].
}
\lref\GaiottoUSA{
  D.~Gaiotto and S.~S.~Razamat,
  ``${\cal N}=1 $ theories of class $ {\cal S}_k $,''
JHEP {\bf 1507}, 073 (2015).
[arXiv:1503.05159 [hep-th]].
}

\lref\qintegOK{
  A.~Okounkov,
  ``(Shifted) Macdonald Polynomials: q-Integral Representation and Combinatorial Formula,''
Compositio Mathematica
June 1998, Volume 112, Issue 2. 
[arXiv:q-alg/9605013].
}
\lref\Lliev{
  A.~Lliev and L.~Manivel,
  ``The Chow ring of the Cayley plane,''
Compositio Mathematica {\bf 141}, Issue 01, 146 (2005).
[arXiv:0306329 [math.AG]].
}
\lref\Gross{
  B.~H.~Gross and N.~R.~Wallach,
  ``On the Hilbert polynomials and Hilbert series of homogeneous projective varieties,''
Advanced Lectures in Mathematics (ALM) {\bf 19}, 253 (2011).
}
\lref\Dolan{
  F.~A.~Dolan,
 ``On Superconformal Characters and Partition Functions in Three Dimensions,''
J.\ Math.\ Phys.\ {\bf 51}, 022301 (2010).
[arXiv:0811.2740 [hep-th]].
}
\lref\Intriligator{
  C.~Cordova, T.~T.~Dumitrescu, and K.~Intriligator,
  ``Deformations of Superconformal Theories,''
[arXiv:1602.01217 [hep-th]].
}

\lref\BeniniNOA{
F.~Benini and A.~Zaffaroni,
``A topologically twisted index for three-dimensional supersymmetric theories,''
JHEP {\bf 1507}, 127 (2015).
[arXiv:1504.03698 [hep-th]].
}

\lref\Closset{
  C.~Closset and I.~Shamir,
  ``The ${\cal N}=1 $ Chiral Multiplet on $T^2 \times S^2$ and Supersymmetric Localization,''
JHEP {\bf 1403}, 040 (2014).
[arXiv:1311.2430 [hep-th]].
}
\lref\Honda{
  M.~Honda and Y.~Yoshida,
  ``Supersymmetric index on $T^2 \times S^2$ and elliptic genus,''
[arXiv:1504.04355 [hep-th]].
}

\Title{\vbox{\baselineskip12pt
}}
{\vbox{
\centerline{Exceptionally simple exceptional models}
\vskip7pt 
\centerline{}
}
}

\centerline{Shlomo S. Razamat and Gabi Zafrir}
\bigskip
\centerline{{\it  Physics Department, Technion, Haifa, Israel 32000}}

\bigskip
\centerline{{\it }}
\vskip.1in \vskip.2in \centerline{\bf Abstract}

We discuss models with no dynamical vector fields in various dimensions which we claim might have exceptional symmetry
on some loci of their parameter space. In particular we construct theories with four  supercharges flowing 
to theories with global symmetry enhancing to $F_4$, $E_6$, and $E_7$.  The main evidence for these claims  is based on 
extracting information about the symmetry properties of the theories from their supersymmetric partition functions. 

\vskip.2in

\noindent

\vfill

\Date{September 2016}

\newsec{Introduction}

Some of the properties of the fixed point, IR or UV, of a general quantum field theory are not obvious from 
a given non-conformal description. For example, the global symmetry at the fixed point might be enhanced in dimension and/or rank. Such symmetry enhancements are often encountered when discussing gauge theories in various dimensions. For example, extended Dynkin diagram shaped ${\cal N}=2$ quiver gauge theories in three dimensions have 
IR fixed points with flavor Lie group corresponding to the Dynkin diagram \IntriligatorEX.  In five dimensions ${\cal N}=1$ $SU(2)$ gauge theories with $N_f<8$ flow in the UV to fixed points with $E_{N_f+1}$ flavor symmetry \SeibergBD.  The enhancement of symmetry  is due to instantons in the latter case and monopoles in the former.  
Moreover in some cases the enhancement of symmetry might only occur on a sub locus of the conformal manifold without obvious explanation due to non-perturbative effects. As an example we mention the enhancement to  $E_7$ of the flavor symmetry of two copies of four dimensional ${\cal N}=1$ $SU(2)$ SQCD with four flavors coupled through a quartic superpotential~\DimoftePD.

In this note we discuss certain  models with four  supercharges constructed from chiral fields and no vector fields in various dimensions. We will present evidence that,  choosing the superpotentials in a careful way, these  models flow to conformal theories, either in the IR or (potentially) in the UV, with conformal manifolds with possible loci having exceptional symmetries.  The superpotential can be constructed in any theory allowing for four supercharges and thus will have this property in two ($(2,2)$ supersymmetry), three (${\cal N}=2$ supersymmetry), or four (${\cal N}=1$ supersymmetry) dimensions. In different dimensions the superpotential will be either relevant or irrelevant leading to fixed points with extended supersymmetry either in the IR or (possibly) UV.  

The arguments in favor of enhancement of global symmetry are based on analysis of partition functions. First we show that the partition functions are invariant under the action of the Weyl group of that symmetry on the parameters, and in case these are indices can be  expanded in characters of the enhanced flavor group. Moreover we will present an argument in three dimensions, generalizing the four dimensional claim \BeemYN\ that in a certain order of the expansion of the index one can extract the number of marginal operators minus the currents. This will let us identify the currents of the enhanced symmetry. The physical interpretation of this result is that when the partition functions are consistent with the enhanced symmetry there is a possibility of a locus of parameter space of the theory at which the symmetry is actually enhanced.  

The note is organized as follows. We will first discuss a simple example of such a  model leading to $SO(8)\times U(1)\times U(1)$, $SO(10)\times U(1)$, and $E_6$ flavor symmetry in section two. We will proceed to deforming the superpotential to obtain a theory with $F_4$ symmetry in section three. In section four we will consider a deformation of two copies of the theory with $E_6$ flavor symmetry leading to a model with $E_7$ symmetry. 
We will discuss several general issues following from our construction in section five. 

\

\newsec{ Model with $E_6$ symmetry}

Let us consider a  model built from $24$ chiral fields. We will organize the fields into six bi-fundamentals of $SU(2)\times SU(2)$. We will have four different $SU(2)$ flavor groups and denote the chiral fields by $Q_1$, $\widetilde Q_1$, $Q_2$, $\widetilde Q_2$, $X$, and $Y$. The superpotential is given by,

\eqn\supsoe{
W_{SO(8)\times U(1)\times U(1)}=Q_1\widetilde Q_1 X+ Q_1 \widetilde Q_2 Y+\widetilde Q_2 Q_2 X+Q_2 \widetilde Q_1 Y\,.
} We can encode this superpotential in the tetrahedral quiver diagram of Fig. 1.

\centerline{\figscale{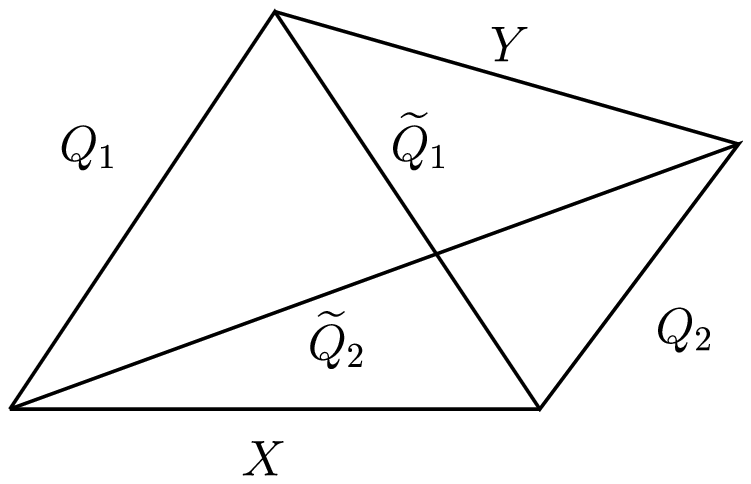}{2.1in}}
\medskip\centerline{\vbox{
\baselineskip12pt\advance\hsize by -1truein
\noindent\footnotefont{\bf Fig.~1:} The superpotential encoded in the quiver. The nodes are $SU(2)$ flavor groups and the lines are bi-fundamental chiral fields. We have a superpotential term for each face of the tetrahedron.}} 

This  model has the manifest symmetry $SU(2)^4\times U(1)_\alpha\times \times U(1)_t$. The four $SU(2)$ symmetries are manifest in the description above and under the $U(1)_t$ the $Q_i$ and  $\widetilde Q_i$ have charge $\frac12$ while $X$ and $Y$ have charge $-1$. Under $U(1)_\alpha$ $Q_i$ have charge $1$ and $\widetilde Q_i$ have charge $-1$ while $X$ and $Y$ have vanishing charge. All fields have R charge $\frac23$. We have also summarized the various charges, including those for fields we shall introduce later in the paper, in table 1. The superpotential is irrelevant in four dimensions, and relevant in lower dimensions. Thus we will think of the model as flowing to an IR fixed point in two and three dimensions while considering a possible flow to a UV completed fixed point in four dimensions.

\centerline{\vbox{\offinterlineskip\tabskip=0pt
\halign{\strut\vrule#
&~#~\hfil\vrule
&~#~\hfil\vrule
&~#~\hfil\vrule
&~#~\hfil\vrule
&~#~\hfil\vrule
&~#~\hfil\vrule
&~#~\hfil\vrule
&~#~\hfil
&\vrule#
\cr
\noalign{\hrule}
&  & $SU(2)_1$ & $SU(2)_2$ & $SU(2)_3$ & $SU(2)_4$ & $U(1)_{\alpha}$ & $U(1)_{t}$ & $U(1)_{R}$& \cr
\noalign{\hrule}
& $Q_1$ & ${\bf 2}$ & ${\bf 2}$ & ${\bf 1}$ & ${\bf 1}$ & $\;\;1$ & $\;\;\frac12$ & $\;\frac23$& \cr
\noalign{\hrule}
& $Q_2$ & ${\bf 1}$ & ${\bf 1}$ & ${\bf 2}$ & ${\bf 2}$ & $\;\;1$ & $\;\;\frac12$ & $\;\frac23$& \cr
\noalign{\hrule}
& $\widetilde Q_1$ & ${\bf 1}$ & ${\bf 2}$ & ${\bf 2}$ & ${\bf 1}$ & $-1$ & $\;\;\frac12$ & $\;\frac23$& \cr
\noalign{\hrule}
& $\widetilde Q_2$ & ${\bf 2}$ & ${\bf 1}$ & ${\bf 1}$ & ${\bf 2}$ & $-1$ & $\;\;\frac12$ & $\;\frac23$& \cr
\noalign{\hrule}
& $X$ & ${\bf 2}$ & ${\bf 1}$ & ${\bf 2}$ & ${\bf 1}$ & $\;\;0$ & $-1$ & $\;\frac23$& \cr
\noalign{\hrule}
& $Y$ & ${\bf 1}$ & ${\bf 2}$ & ${\bf 1}$ & ${\bf 2}$ & $\;\;0$ & $-1$ & $\;\frac23$& \cr
\noalign{\hrule}
\noalign{\hrule}
& $Z_+$ & ${\bf 1}$ & ${\bf 1}$ & ${\bf 1}$ & ${\bf 1}$ & $\;\;2$ & $-1$ & $\;\frac23$& \cr
\noalign{\hrule}
& $Z_-$ & ${\bf 1}$ & ${\bf 1}$ & ${\bf 1}$ & ${\bf 1}$ & $-2$ & $-1$ & $\;\frac23$& \cr
\noalign{\hrule}
\noalign{\hrule}
& $\Lambda$ & ${\bf 1}$ & ${\bf 1}$ & ${\bf 1}$ & ${\bf 1}$ & $\;\;0$ & $\;\;2$ & $\;\frac23$& \cr
 }\hrule}}
\medskip\centerline{\vbox{
\baselineskip12pt\advance\hsize by -1truein
\noindent\footnotefont{\bf Table~1:} The fields appearing in the models discussed in this section, together with their charges under the various global symmetries.}} 

\

First, we claim that the $SU(2)^4$ symmetry here enhances to $SO(8)$.  We can check this by studying different supersymmetric partition functions in various dimensions. For example, in four dimensions the index is given by,\foot{For notations and definitions of supersymmetric partition functions the reader can consult~\PestunZXK .}

\eqn\indf{\eqalign{
1+({\bf 8_s}t^{\frac12}\alpha^{-1} +{\bf 8_v}t^{\frac12}\alpha^{}+{\bf 8_c}t^{-1})(pq)^{\frac13}+\cdots+
(4-{\bf 28}-1-1+{\bf 350}+\cdots)pq+\cdots
}
}  The interpretation \BeemYN\ of this result is that if there is a  UV fixed point for which this is the index,  it has an $SO(8)\times U(1)\times U(1)$ flavor symmetry ($-{\bf 28}-1-1$ terms in the order $p q$ of the index corresponding to the conserved currents), and it has a conformal manifold of dimension  $4$ preserving this symmetry. Symmetry properties of the $4d$ index also give the symmetry of the $\S^3$ partition function in  three dimensions. In three dimensions this theory flows to a CFT in the IR.  In the next section we will generalize the arguments of \BeemYN\ to three dimensions, and show that also the index in three dimensions exhibits this symmetry.
 We can write down  other partition functions in other dimensions exhibiting the symmetry (elliptic genus, spheres, indices). The details of the physics will depend on the dimension but the symmetry will remain.

\

We can slightly complicate the model by adding more fields. For example, we can add the fields $Z_\pm$ which 
are singlets under $SU(2)^4$, have $U(1)_t$ charge $-1$, $U(1)_\alpha$ charges $\pm2$, and R charge $\frac23$. 
We couple these fields as,

\eqn\supt{
W_{SO(10)\times U(1)}=W_{SO(8)\times U(1)\times U(1)}+Z_+(\epsilon\cdot \widetilde Q_1^2+\epsilon\cdot \widetilde Q_2^2)+Z_-(\epsilon\cdot  Q_1^2+\epsilon\cdot  Q_2^2)\,.
} This theory has symmetry $SO(10)\times U(1)_t$, where $SO(8)\times U(1)_\alpha$ enhances to $SO(10)$. Giving the example of the index in four dimensions we obtain,

\eqn\sott{
1+({\bf 10}t^{-1} +{\bf 16}t^{\frac12})(pq)^{\frac
13}+\cdots+
(-{\bf 45}-1+{\bf\overline{1050} }+\cdots)pq+\cdots
} Here we deduce that we have $SO(10)\times U(1)$ symmetry (the $-1-{\bf 45}$ giving the currents) and that there is no marginal operator preserving this symmetry.

\

We can add another field to enhance the symmetry farther. We add a single field $\Lambda$ which is charged under $U(1)_t$ with charge $2$ and has R charge $\frac23$, and is a singlet under all the other symmetries. The superpotential is,

\eqn\supeg{
W_{E_6}=W_{SO(10)\times U(1)}+\Lambda (X^2+Y^2+Z_-Z_+)\,.
} This theory has an $E_6$ flavor symmetry. Again in four dimensions the index is,

\eqn\esin{
1+{\bf 27} (p q)^{\frac13}+{\bf \overline{351}'}(p q)^{\frac23}+(-{\bf 78}+{\bf 3003}) pq+\cdots\,.
} Like in the previous cases, this result suggests that if there is a UV fixed point, for which this is the index, it has an $E_6$ global symmetry. One can also consider the analogue theory in three or two spacetime dimensions. Particularly, in section $2.2$ we shall examine the $3d$ index of the analogous three dimensional theory and argue that we can derive a similar result also for this case. However now the theory is expected to flow to an IR fixed point and the index can be readily interpreted as its supercoformal index.

The fact that the superpotential \supeg\ gives rise to $E_6$ symmetry is not surprising.  One can realize this symmetry as the group of transformations fixing the determinant of a three by three hermitian matrix built from octonions. This determinant gives  the polynomial $W_{E_6}$ with very specific numerical coefficients. Since the supersymmetric partition functions are insensitive to such parameters we allow ourselves to be agnostic about them in our discussion. 

\

\subsec{The moduli space}

The  model we presented has a moduli space spanned by the vacuum expectation values of the scalars in the chiral fields modulo the superpotential constraints. In this subsection we try to identify this space. The physics data describes it as an algebraic variety of $\C^{27}$ defined by $27$ quadratic equations. As a first step we note that the equations are homogeneous so the moduli space must be a complex cone over another space $B$.

This structure of the superpotential leads to two interesting features. First, there should be a conical singularity at the origin. This is expected as there are massless fields there. Second, there is a natural $U(1)$ action on the cone which we identify as the $U(1)_R$ symmetry of the theory. Indeed all fields have the same R-charge which agrees with the $U(1)$ action on the cone.

So now we need to identify the space $B$ which the equations define as an algebraic variety of $\C\P^{26}$. We propose that this space is the complex Cayley plane which is a $16$ dimensional complex manifold. This space can indeed be defined as an algebraic variety of $\C\P^{26}$ via $27$ quadratic equations \Lliev. Alternatively it can be defined as the symmetric space $E_6/(SO(10)\times U(1))$. This definition manifests its $E_6$ isometry.

We can also provide additional evidence for this identification. First the Hilbert series for the complex Cayley plane was calculated in \Gross. The first few terms of their results suggest the space is spanned by functions in the ${\bf 27}$ of $E_6$ subject to the condition that the ${\bf \overline{27}}$ does not appear in their symmetric product. This agrees with the result we observe from the index.      

We can also try to infer the dimension of the manifold from the equations. Say we choose a non-singular point on the manifold and expand the equations around this point. We can then linearize the equations and solve the resulting linear system. The dimension of the solution space is then the dimension of the manifold. Of course this only works if we choose a non-singular point. As the Cayley plane is a symmetric space, if the moduli space is as we proposed, any point save the origin will do.  Say we take all fields to be zero except: $\Lambda,  (Q_1)_{22},  (Q_2)_{22}, (\widetilde Q_1)_{22}$ and $(\widetilde Q_2)_{22}$ (we regard the bifundamental chirals as matrices and use the subscript as an entry in the matrix). It is easy to check that this is a solution. We then expanded around this solution and found that there is indeed a $17$ dimensional solution space in accordance with our picture of the moduli space\foot{Note that this requires some tuning of the constants appearing in the superpotential.}.  

\

\subsec{$3d$ supersymmetric index}

We can also look at partition functions in other dimensions, notably the $3d$ supersymmetric index and $2d$ elliptic genus. We shall show that they also can be expanded in characters of $E_6$. We shall start with the $3d$ supersymmetric index as for the $3d$ case the $E_6$ model leads to a conformal fixed point in the IR. For the $3d$ theory the supersymmetric index is given by,

\eqn\esin{
1+{\bf 27} x^{\frac23}+{\bf\overline{351}'} x^{\frac43}+(-{\bf 78}+{\bf 3003}) x^2+\cdots\,.
} This is very similar in structure to the $4d$ index. Again all states fall in characters of $E_6$.

An interesting question is whether we can identify the superconformal multiplets contributing to the index similarly to the results \BeemYN\ we stated for the $4d$ index. For this we consider the possible short multiplets and their contribution to the index. The shortening conditions for various $3d$ superconformal algebras were extensively discussed in \Dolan. A more concise summary can be found in \Intriligator, and we shall employ their notations for the various short multiplets.

The $3d$ $\cal{N}$$=2$ superconformal algebra contains $2$ fermionic supersymmetry generators denoted as $\cal{Q}$ and $\bar{\cal{Q}}$. The superconformal multiplet is then generated by acting with them, and with the translation generators $P_{\mu}$, on a superconformal primary\foot{These are states annihilated by the generators of special conformal transformations $K_{\mu}$, and their fermionic partners $\cal{S}$ and $\bar{\cal{S}}$.}. Short representations are those for which the superconformal primary is annihilated by some combination of $\cal{Q}$s and $\bar{\cal{Q}}$s. Due to the superconformal algebra this necessarily fixes the dimension of the superconformal primary in term of its R-charge and angular momentum. 

\

\

\vbox{\offinterlineskip\tabskip=0pt
\halign{\strut\vrule#
&~#~\hfil\vrule
&~#~\hfil\vrule
&~#~\hfil
&\vrule#
\cr
\noalign{\hrule}
&  & Shortening conditions & Index& \cr
\noalign{\hrule}
 & $A_1 \bar{L}$ & $\epsilon^{a b}$ $\cal{Q}$$_{a} |SCP\rangle_b = 0$, $\Delta=j-r+1$, $j\geq \frac12$, $r<0$ & $0$ & \cr
 & $A_2 \bar{L}$ & $ (\cal{Q})$$^2 |SCP\rangle = 0$, $\Delta=-r+1$, $j=0$, $r<0$ & $0$ & \cr
 & $L \bar{A}_1$ & $\epsilon^{a b} $$\bar{\cal{Q}}$$_{a} |SCP\rangle_b = 0$, $\Delta=j+r+1$, $j\geq \frac12$, $r>0$ & $I(2+r+2j,2j+1)$ & \cr 
 & $L \bar{A}_2$ & $ (\bar{\cal{Q}})$$^2 |SCP\rangle = 0$, $\Delta=r+1$, $j=0$, $r>0$ & $I(2+r,1)$ &  \cr 
 & $B_1 \bar{L}$ & $\cal{Q}$$_{a} |SCP\rangle = 0$, $\Delta=-r$, $j=0$, $r<-\frac12$ & $0$ & \cr
 & $L \bar{B}_1$ & $\bar{\cal{Q}}$$_{a} |SCP\rangle = 0$, $\Delta=r$, $j=0$, $r>\frac12$ & $I(r,0)$ &  \cr
 & $A_1 \bar{A}_1$ & $\epsilon^{a b}$$ \cal{Q}$$_{a} |SCP\rangle_b = 0$ and  $\epsilon^{a b}$$ \bar{\cal{Q}}$$_{a} |SCP\rangle_b = 0$, $\Delta=j+1$, $j\geq \frac12$, $r=0$ & $I(2j+2,2j+1)$ & \cr
 & $A_2 \bar{A}_2$ & $ (\cal{Q})$$^2 |SCP\rangle = 0$ and  $ (\bar{\cal{Q}})$$^2 |SCP\rangle = 0$, $\Delta=1$, $j=r=0$ & $I(2,1)$ & \cr
 & $B_1 \bar{A}_2$ & $\cal{Q}$$_{a} |SCP\rangle = 0$ and  $ (\bar{\cal{Q}})$$^2 |SCP\rangle = 0$, $\Delta=\frac12$, $j=0$, $r=-\frac12$ & $I(\frac32,1)$ & \cr
 & $A_2 \bar{B}_1$ & $\bar{\cal{Q}}$$_{a} |SCP\rangle = 0$ and  $ (\cal{Q})$$^2 |SCP\rangle = 0$, $\Delta=\frac12$, $j=0$, $r=\frac12$ & $I(\frac12,0)$ & \cr
 }\hrule}
\medskip\centerline{\vbox{
\baselineskip12pt\advance\hsize by -1truein
\noindent\footnotefont{\bf Table~2:} The shortening conditions and index contributions for the various short multiplets, where we have adopted the notations of \Intriligator\  in the naming of the various short multiplets. We have also used $|SCP\rangle$ for the state associated with the superconformal primary, $\Delta$ for its conformal dimension, $r$ for its R-charge and $j$ for its angular momentum.}}  

\

\

In table 2 we have summarized the various short representations, shortening conditions and their contribution to the index. For the index contribution we have defined,

\eqn\esin{
I(l,s)=(-1)^s\frac{x^l}{1-x^2}.
} 

We now study what multiplets can contribute to the $3d$ index at order $x^l$ for $l \leq 2$. From table 2 we see that the only multiplets that can contribute are: $A_2 \bar{B_1}$, $B_1 \bar{A_2}$, $L \bar{B}_1$ and $A_2 \bar{A}_2$. The multiplets $A_2 \bar{B_1}$ and $B_1 \bar{A_2}$ are free fields and indeed their combination is the free chiral multiplet. The $L \bar{B}_1$ type multiplets are chiral fields and thus their contributions are the relevant operators for $l<2$ and marginal operators for $l=2$. The $A_2 \bar{A}_2$ is the conserved current multiplet. 

From this we see that, similarly to $4d$, the $x^2$ order is the marginal operators minus the conserved currents. Particularly for the $E_6$  model we indeed see a negative contribution, at order $x^2$, in the adjoint of $E_6$. This supports our claim that this model has an IR fixed point with $E_6$ global symmetry somewhere on its conformal manifold. 

\

\

\subsec{Other partition functions}

We can also evaluate the $2d$ elliptic genus, which is given by,

\eqn\esin{
PE\biggl[\frac{y^{\frac13} {\bf 27}}{1-q} + \frac{q^{-1} y^{-\frac13} {\bf \bar{27}}}{1-q} - \frac{y^{-\frac23} {\bf 27}}{1-q} - \frac{q^{-1} y^{\frac23} {\bf \bar{27}}}{1-q}\biggr],
} where $PE$ stands for plethystic exponential. The structure again has some similarities with the $4d$ and $3d$ indices though it contains more terms. Particularly it can be cast in characters of $E_6$. 

We can also calculate other $4d$ partition functions. For instance, the lens space index and the $\S^2 \times T^2$ partition function. The latter is hindered by the fact that it (see \refs{\Closset,\Honda,\BeniniNOA}) requires integer R-charges. We can try to correct this by mixing the $U(1)_R$ symmetry with $U(1)_t$ which in the $\S^2 \times T^2$ partition function formalism is associated with adding magnetic flux on $\S^2$. Unfortunately adding the magnetic flux breaks $E_6$ down to its $SO(10)\times U(1)$ subgroup. The resulting partition function depends on the choices of the magnetic fluxes but can be expressed in characters of $SO(10)\times U(1)$.  

The lens space index is quite similar to the $4d$ index, but with more terms, and it in general depends on the chosen lens space, $\S^3/Z_k$. A novelty in this index is that one can accommodate a non-trivial $\Z_k$ holonomy on $\S^3/\Z_k$ for flavor symmetries. For the  model we consider without holonomies, the lens space index reads,

\eqn\esin{
PE\biggl[\frac{{\bf 27} (p q)^{\frac13} - \overline{{\bf 27}} (p q)^{\frac23}}{(1-q)(1-p)} F_k(p,q)\biggr]. 
} This differs from the $4d$ index by the factor of $F_k(p,q)$ whose exact form is given in \BeniniNC. The expression is inherently written in characters of $E_6$. Adding holonomies under a collection of $U(1)$'s will change the factor $F_k(p,q)$ for each chiral field based on its charges under these symmetries. Naturally this will break $E_6$. Still we retain the action of the Weyl group which implies that different holonomies, related by the action of the Weyl group, should have the same index. This again is manifest in the expression as the chiral fields sit in characters of $E_6$ which ensures they are properly transformed under the action of the Weyl group. 

\

\newsec{Model with $F_4$ symmetry}

Let's return to the $SO(10)\times U(1)$  model, and consider reducing the symmetry by enlarging the superpotential. Specifically, we consider breaking $U(1)_t$ and $U(1)_{\alpha}$ while preserving $U(1)_R$ and the four $SU(2)$ groups. Adding all terms compatible with these requirements gives the superpotential,

\eqn\supeg{
W_{F_4}=W_{SO(10)\times U(1)}+(Z_- + Z_+)(X^2+Y^2+\epsilon\cdot \widetilde Q_1^2+\epsilon\cdot \widetilde Q_2^2+\epsilon\cdot  Q_1^2+\epsilon\cdot  Q_2^2)+Z^3_- +Z^2_- Z_+ +Z^2_+ Z_- + Z^3_+\,.
}  This theory has an $F_4$ flavor symmetry. Specifically, we consider the three dimensional  model for which the $3d$ index is,

\eqn\esin{
1+{\bf 26} x^{\frac23}+({\bf 324}+ 1)x^{\frac43}+(-{\bf 52}+{\bf 2652}) x^2+\cdots\,.
} We interpret this as the IR fixed point of this  model having a conformal manifold with a point with enhanced symmetry which is $F_4$ in this case. The four dimensional index, relevant for the $4d$  model, also has a similar structure but with $x^2$ replaced by $p q$. Since the superpotential is irrelevant in $4d$, this  model is only interesting if there is a UV completed fixed point. 

We can again inquire about the moduli space. The structure of the equations is quite similar so we again expect the moduli space to be a complex cone over another space $B$. The space $B$ can be described as an algebraic variety of $\C\P^{25}$ by $26$ quadratic equations. We also expect $B$ to have an $F_4$ isometry. A natural guess is that $B$ is a symmetric space similarly to the $E_6$ case. This is reasonable as given a solution to the equations we can generate more solutions by acting with the $F_4$ global symmetry. Assuming this covers all solutions, the resulting space is a symmetric space given by $F_4$ moded by the symmetry keeping the solution fixed.

There are two compact symmetric spaces with $F_4$ isometry: $F_4/SO(9)$ (the real Cayley plane) and $F_4/(SU(2)\times USp(6))$. The first is $8$ complex dimensional space and the second is $14$ complex dimensional space. We next analyze the equations linearized around the solution where the only non-vanishing fields are: $(Q_1)_{22},  (Q_2)_{22}, (\widetilde Q_1)_{22}$ and $(\widetilde Q_2)_{22}$. We find a $15$ dimensional solution space\foot{This requires some tuning of the constants in the superpotential.}. This is consistent with the moduli space being a complex cone over the symmetric space $F_4/(SU(2)\times USp(6))$.

\

\newsec{Model with $E_7$ symmetry}

We can use the $E_6$  model to generate a model with $E_7$ global symmetry. To do this we take $56$ chiral multiplets and split them into two copies of the $27$ chiral fields in the $E_6$  model and two additional chiral fields $P_+$ and $P_-$. The fields interact through the superpotential, 

\eqn\supeg{
W_{E_7}=P_+ W^1_{E_6} + P_- W^2_{E_6} + W_{int} \,,
}  where we use $W^1_{E_6}$ and $W^2_{E_6}$ for the superpotential of the $E_6$  model involving chiral fields from just one of the two copies,  these being the first or second copy respectively. We use $W_{int}$ for the most general quartic superpotential involving only the combinations $P_+ P_-$ and products of fields in the first copy with its image in the second copy. 

The classical flavor symmetry is $SU(2)^4\times U(1)_t \times U(1)_{\alpha} \times U(1)_p$. Fields belonging to one copy of the $27$ transform as before under $SU(2)^4\times U(1)_t \times U(1)_{\alpha}$ while the other copy transforms as the complex conjugate. Under $U(1)_p$ copy one has charge $-1$ while copy two has charge $1$. The fields $P_+$ and $P_-$ are singlets under $SU(2)^4\times U(1)_t \times U(1)_{\alpha}$ and carry charge $3$ and $-3$ under $U(1)_p$, respectively. 

The theory also has a $U(1)_R$ symmetry where now the R-charge of all the fields is $\frac12$. The superpotential is now quartic so it is irrelevant in four dimensions, marginally irrelevant in three dimensions and relevant in two dimensions. 

We claim that this theory has $E_7$ global symmetry. Again in four dimensions the index is,

\eqn\esin{
1+{\bf 56} (p q)^{\frac14}+({\bf 1463}+ {\bf 133})(p q)^{\frac12}+({\bf 24320}+{\bf 6480})(p q)^{\frac34}+(-{\bf 133}+{\bf 293930}+ {\bf 150822}+ {\bf 7371}) pq+\cdots\,.
} The three dimensional index also has a similar structure but with $p q$ replaced by $x^2$. If either the $3d$ or the $4d$ models possess a UV completed fixed point, then the indices suggest it should have an $E_7$ global symmetry. 

The analogue two dimensional  model is expected to flow to an IR fixed point, the elliptic genus of which, can be cast in characters of $E_7$. Therefore one may also expect this IR fixed point to have an $E_7$ global symmetry at a point on its conformal manifold.

\

\newsec{General properties}

Finally, we wish to discuss some general properties that emerge from our construction. Specifically we seek to summarize the salient features of our construction in a way that facilitates generalizations to other systems. In general we have a collection of chiral fields that we choose to form a representation $R$ of a chosen group $G$, where for simplicity we consider only a single representation of $G$. The chiral fields carry charges under the classical symmetry so that they correctly form the representation $R$ of $G$. 

This generally requires a superpotential to force all fields to carry the desired charges and eliminates additional symmetries. We shall limit ourselves to theories with an $R$-symmetry as in these cases we can preform more stringent tests using the superconformal index. Furthermore, as by assumption all chiral fields form a single representation $R$ of $G$, they must have the same $R$-charge. The results of these two conditions is that the superpotential must be a polynomial in the fields of degree $r$. One obstruction for this construction is that one must be able to find the desired superpotential. We can formulate some necessary conditions using group theory.

 First, group theory gives a limitation on the possible values of $r$. The chiral ring of the theory is made from the symmetric products of the chiral fields and so is in $G$ representations appearing in such products. The superotential constraints eliminate chiral ring elements made from the $r-1$ symmetric product of the basic chiral fields, and carry charges in the conjugate representation to $R$. Thus consistency necessitates that the representation $\bar{R}$ must appear in $\otimes^{r-1}_{Sym} R$. This in turn constrains $r$.

For example, for $E_6$ we have $\otimes^{2}_{Sym} {\bf 27} = {\bf\overline{351}'}+{\bf \overline{27}}$ so the minimal possible value of $r$ is $3$. Likewise for $F_4$, $\otimes^{2}_{Sym} {\bf 26} = {\bf 324}+{\bf 26} + {\bf 1}$ so the minimal non-trivial value of $r$ is again $3$. However for $E_7$ we have $\otimes^{2}_{Sym} {\bf 56} = {\bf 1463}+ {\bf 133}$ so a cubic superpotential is not possible. Yet $\otimes^{3}_{Sym} {\bf 56} \supset {\bf 56}$ so the minimal non-trivial value of $r$ in this case is $4$. 

An additional condition can then be given using the $4d$ supersymmetric index (or as we have seen the three dimensional one). This receives contributions from the chiral fields modulo the superpotential constraints. A nice feature of the $4d$ index is that the $p q$ order receives contributions only from marginal operators, which contribute positively, and conserved currents, which contribute negatively. Therefore we can look at the negative terms in the $p q$ order and see whether or note we indeed get the adjoint, and only the adjoint representation of $G$. In fact it is straightforward to write the contribution for the $p q$ order to be: $\otimes^{r}_{Sym} R - R \otimes \bar{R}$. This essentially reduces the problem to group theory: which representation appearing in the direct product $R \otimes \bar{R}$ do not appear in $\otimes^{r}_{Sym} R$. For example, in the $E_6$, $F_4$ and $E_7$ theories the answer to this is indeed only the adjoint representation. 

\

\subsec{Example: $G_2$}

As an illustrating example let's consider the exceptional group $G_2$ and its $\bf{7}$ dimensional representation. It is convenient to form the chiral fields in representations of the $SU(3)$ maximal subgroup of $G_2$. Under it the $\bf{7}$ of $G_2$ decomposes as $\bf{1} + \bf{3} + \bar{\bf{3}}$ so we shall use $3$ chiral fields $F$ in the $\bf{3}$ of the classical $SU(3)$, $3$ chiral fields $\bar{F}$ in the $\bar{\bf{3}}$ and a singlet X. Next we need to find a superpotential that limits the fields to these charges. However we shall now argue that this is not possible. 

The superpotential must be $SU(3)$ invariant and so must be made from the meson $F \bar{F}$. Note that baryonic products vanish as the fields are bosonic. This implies that the minimal non-trivial order for the superpotential is $4$, and also that there is an additional $U(1)$ under which $F$ and $\bar{F}$ carry opposite charges that we cannot eliminate. The superpotential that we can add has the form,

\eqn\supeg{
 W = (F \bar{F})^2 + F \bar{F} X^2 + X^4 \,.
} This leads to a classical $U(1)\times SU(3)$ global symmetry under which the fields are charged as: $\bf{1}^0 + \bf{3}^{1} + \bar{\bf{3}}^{-1}$. These in fact form the $\bf{7}$ of $SO(7)$ under its $U(1)\times SU(3)$ subgroup. So we conclude that we cannot build a $G_2$  model. Attempting to build one leads to model with $SO(7)$ global symmetry. 

We can also see all these statements materialize just from group theory analysis. First note that $\otimes^{2}_{Sym} {\bf 7} = {\bf 27} + {\bf 1}$ and $\otimes^{3}_{Sym} {\bf 7} = {\bf 77} + {\bf 7}$ so indeed the minimal non-trivial order of the superpotential is $4$. Next we look at the conserved currents given by the terms in the product ${\bf 7} \otimes {\bf 7}$ that are not contained in $\otimes^{4}_{Sym} {\bf 7}$. Doing the group theory we find these to be the adjoint $\bf{14}$ of $G_2$ and the $\bf{7}$. Thus we see that there are additional conserved currents, which in fact form the adjoint of $SO(7)$ signaling that such a theory must have a larger global symmetry. So the group theory analysis supports the previous claim that there is no analogous  model wth $G_2$ as its global symmetry.

\

\

\

\noindent {\bf Acknowledgments:}
We would like to thank Nathan Seiberg, Brian Willett and Amos Yarom for useful comments and discussions. GZ is supported in part by the Israel Science Foundation under grant no. 352/13, and by the German-Israeli Foundation for Scientific Research and Development under grant no. 1156-124.7/2011.  SSR is  a Jacques Lewiner Career Advancement Chair fellow. This research was also supported by Israel Science Foundation under grant no. 1696/15 and by I-CORE  Program of the Planning and Budgeting Committee.

\listrefs
\end